\documentclass[,twocolumn,preprintnumbers,amsmath,amssymb,superscriptaddress]{revtex4}

\usepackage{graphicx}% Include figure files
\usepackage{dcolumn}% Align table columns on decimal point
\usepackage{bm}% bold math

\begin{document}

\title{Linear and Nonlinear Optical Constants of BiFeO$_3$}

\author{Amit Kumar}
\affiliation{Department of Materials Science and Engineering, Pennsylvania State University, MRL Bldg., University Park, Pennsylvania 16802, USA}

\author{Ram C. Rai}
\affiliation{ Department of Chemistry, University of Tennessee, Knoxville, TN 37996,USA }

\author{Nikolas J. Podraza}
\affiliation{Department of Materials Science and Engineering, Pennsylvania State University, MRL Bldg., University Park, Pennsylvania 16802, USA}

\author{Sava Denev}
\affiliation{Department of Materials Science and Engineering, Pennsylvania State University, MRL Bldg., University Park, Pennsylvania 16802, USA}

\author{Mariola Ramirez}
\affiliation{Department of Materials Science and Engineering, Pennsylvania State University, MRL Bldg., University Park, Pennsylvania 16802, USA}

\author{Ying-Hao Chu}
\affiliation{Department of Materials Science and Engineering, University of California, Berkeley, Hearst Mining Building, Berkeley, California 94720, USA}
 
\author{Jon Ihlefeld}
\affiliation{Department of Materials Science and Engineering, Pennsylvania State University, MRL Bldg., University Park, Pennsylvania 16802, USA}

\author{T. Heeg}
\affiliation{Institute of Bio- and Nano-Systems (IBN1-IT), Research Centre J$\ddot{u}$lich, D-52425 J$\ddot{u}$lich, Germany}

\author{J. Schubert}
\affiliation{Institute of Bio- and Nano-Systems (IBN1-IT), Research Centre J$\ddot{u}$lich, D-52425 J$\ddot{u}$lich, Germany}

\author{Darrell G. Schlom}
\affiliation{Department of Materials Science and Engineering, Pennsylvania State University, MRL Bldg., University Park, Pennsylvania 16802, USA}
 
\author{J. Orenstein}
\affiliation{Department of Physics, University of California, Berkeley, California 94720, USA}

\author{R. Ramesh}
\affiliation{Department of Materials Science and Engineering, University of California, Berkeley, Hearst Mining Building, Berkeley, California 94720, USA}

\author{Robert W. Collins}
\affiliation{Department of Physics and Astronomy, University of Toledo, Toledo, OH, 43606, USA}

\author{Janice L. Musfeldt }
\affiliation{ Department of Chemistry, University of Tennessee, Knoxville, TN 37996,USA }

\author{Venkatraman Gopalan}
\affiliation{Department of Materials Science and Engineering, Pennsylvania State University, MRL Bldg., University Park, Pennsylvania 16802, USA}

\begin{abstract}
Using spectroscopic ellipsometry,  the refractive index and absorption versus wavelength of the ferroelectric antiferromagnet Bismuth Ferrite, BiFeO$_3$ is reported. The material has a direct  band-gap at 442 nm wavelength (2.81 eV). Using optical second harmonic generation, the nonlinear optical coefficients  were determined to be  $d_{15}/d_{22} = 0.20\pm0.01$, $d_{31}/d_{22} = 0.35\pm0.02$, $d_{33}/d_{22} = -11.4\pm0.20$ and $\left|d_{22}\right|$ = 298.4$\pm$6.1 pm/V at a fundamental wavelength of 800 nm.
\end{abstract}
\maketitle
BiFeO$_3$ (BFO) is an antiferromagnetic, ferroelectric  with  Neel temperature T$_N$ = 643 K, and ferroelectric Curie temperature T$_C$ = 1103 K.\cite{1,2,3} It is presently one of the most studied multiferroic materials due to its large ferroelectric polarization of $\sim$ 100 $\mu$C/cm$^2$ in thin films, and the possibility of  coupling between magnetic and ferroelectric order parameters, thus enabling manipulation of one through the other.\cite{1} Linear and nonlinear optical spectroscopy tools are ideally suited to study such coupling.\cite{4} While the mean refractive index  for bulk single crystal BiFeO$_3$ has been previously investigated,\cite{5} the optical constants of thin films have not been presented thus far. Also, an indirect gap at 673 nm (1.84 eV) was reported before\cite{6}, which is shown here to be an absorption onset potentially due to a joint density of states effect and not associated with phonon participation. In our analysis, the material appears to have a direct gap with a bandedge at 442 nm instead. No studies of nonlinear optical coefficients of BiFeO$_3$, in any form, exist. In this letter, we measure large second order optical nonlinearities in  BFO.
\begin{figure}[h!t]
 \centering
 \includegraphics[width=0.48 \textwidth,height=0.38 \textheight,angle=0]{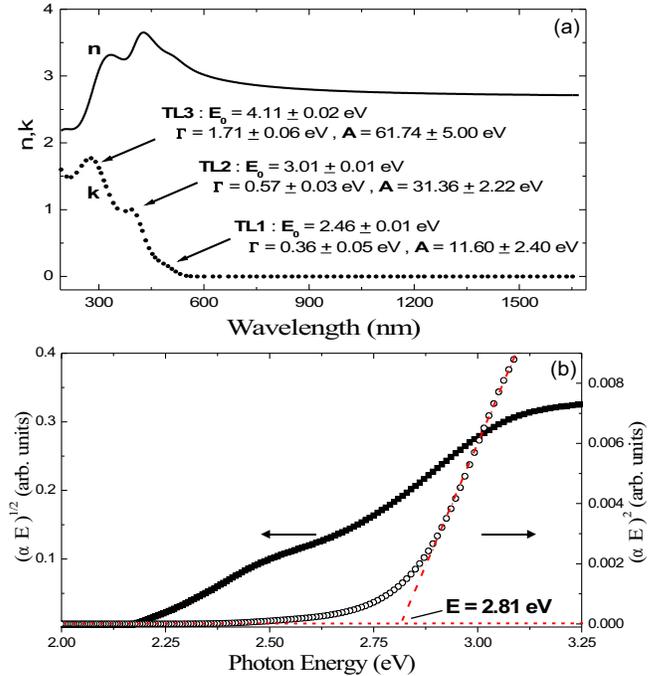}
 \caption{\label{fig:1}(a)Index of refraction (n) and extinction coefficient (k) for BiFeO$_3$ deposited on (111) SrTiO$_3$ over a spectral range from 190 to 1670 nm. Also shown are the dielectric function parameters for the three Tauc-Lorentz oscillators (TL 1, TL 2, TL 3) within the spectral range.  The parameterization also includes an additional oscillator outside the spectral range with E$_0$ = 7.29 $\pm$ 0.06 eV, $\Gamma$ = 4.28 $\pm$ 0.34 eV and A = 51.19 $\pm$ 2.78 eV. (b) Plot of ($\alpha$E)$^{1/2}$ and ($\alpha$E)$^2$ vs. photon energy E where a linear extrapolation of ($\alpha$E)$^2$ to 0 suggests a direct bandgap at 2.81 eV.}
\end{figure}
 
 Epitaxial and phase-pure BFO thin films were synthesized by pulsed-laser deposition (PLD) as well as molecular-beam epitaxy (MBE) on (111) SrTiO$_{3}$ (STO) substrates.\cite{1,7} The films studied here are epitaxial with orientation relationship BFO(0001)//STO(111)and [2$\bar{1}\bar{1}$0]BFO//[1$\bar{1}$0] STO. We note specifically that unlike many epitaxial thin films, these films do not have any additional structural variants, including any rotational variants within the film growth plane. Thus, these (0001) oriented films have nearly single crystalline perfection, with three well-defined crystallographic \textit{x}-[2$\bar{1}\bar{1}$0], and \textit{y}-[1$\bar{1}$00] axes within the film plane, and the \textit{z}-[0001] axis normal to the plane. The three \textit{y}-\textit{z} mirror planes in the 3$m$ point group symmetry for BFO are thus well defined and allow us to extract nonlinear coefficients precisely without ambiguity.  Typical film stoichiometry, as determined by Rutherford backscattering spectrometry (RBS), was stoichiometric within  ~3 \% error of the measurement (Bi:Fe = 0.98-0.99:1). There were no amorphous or secondary phases as confirmed by transmission electron microscopy. 
\begin{figure}
 \centering
\includegraphics[width=0.49 \textwidth,angle=0]{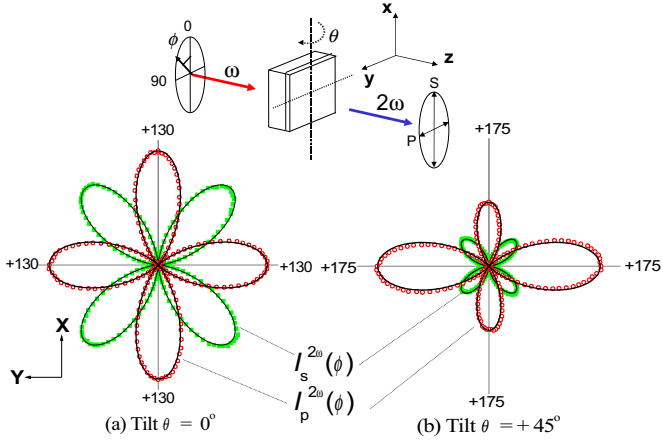}
 \caption{\label{fig:2} Variation of $\textit{p}$ and $\textit{s}$ polarized SHG intensity with incident polarization angle for a BFO//STO(111) film with \textit{x}-axis perpendicular to the plane of incidence in  (a) Normal incidence $\theta = 0^\circ$ and (b) $\theta = 45^\circ$  tilt.}
\end{figure}

 Ellipsometric spectra in ($\Delta,\Psi$) were collected ex situ for a BiFeO$_3$ film prepared by MBE on (111) SrTiO$_3$ at $\theta_i = 55^\circ$ and $70^\circ$ angles of incidence using a variable-angle rotating-compensator multichannel spectroscopic ellipsometer\cite{8} with a spectral range from 190 to 1670 nm. The optical properties (n, k) shown in Fig. \ref{fig:1}(a) and the corresponding dielectric function spectra ($\varepsilon_1$,$\varepsilon_2$) are extracted by using a least squares regression analysis and a weighted root mean square error\cite{9}, to fit the ellipsometric spectra to a four-medium optical model consisting of a semi-infinite STO substrate / bulk film / surface roughness / air ambient structure. The free parameters correspond to the bulk and surface roughness thicknesses of the film and a parameterization of the BiFeO$_3$ dielectric function. The dielectric function parameterization of BiFeO$_3$ consists of four Tauc-Lorentz oscillators \cite{10} sharing a common band gap and a constant additive term to $\varepsilon_1$ denoted by $\varepsilon_\infty$ (equal to 1 for this model). The parameters corresponding to each oscillator include an oscillator amplitude A, broadening parameter $\Gamma$, resonance energy $E_0$, and a Tauc gap $E_g$ common to all oscillators. The optical properties of the surface roughness layer are represented by a Bruggeman effective medium approximation consisting of a 0.50 bulk film / 0.50 void mixture.\cite{11} This model yields the common Tauc gap E$_g$ = 2.15$\pm$0.06 eV, bulk thickness d$_b$= 468.93$\pm$0.78 \AA, and surface roughness thickness  d$_s$= 75.39$\pm$0.4 \AA. 

  We note that though the Tauc gap at 2.15 eV (577 nm) represents the onset of absorption, it is \textit{not} the indirect gap, as claimed in literature.\cite{6} A plot of  $\alpha^2$E$^2$ vs. photon energy E ($\alpha$ = 4$\pi$k/$\lambda$) and the linear extrapolation to $\alpha^2$ E$^2$ = 0 indicates a direct gap at 2.81 eV (442 nm) as shown in Fig. \ref{fig:1}(b). This value is in good agreement with that obtained from more recent optical measurements.\cite{12} Band gap measurements on  different MBE-grown BiFeO$_3$ films grown on (001) and (111) SrTiO$_3$ substrates revealed a direct band gap in all cases with E$_g$ = 2.77$\pm$0.04 eV. The presence of two distinct slopes in $(\alpha$E$)^{1/2}$ vs. E characteristic of an indirect band gap \cite{13} is not observed. We obtain the linear complex indices from this model to be $\tilde{N}_f^\omega$=2.836+0$\textit{i}$ and $\tilde{N}_f^{2\omega}$=3.444+0.981$\textit{i}$ for corresponding wavelengths of 800 and 400 nm, respectively. It should be noted that although BiFeO$_3$ is uniaxially anisotropic, only the optical properties of the ordinary index of refraction have been obtained for this film.\cite{14} 

The crystal symmetry of epitaxial BFO(111) films has been shown to be point group 3$m$ using optical SHG and diffraction techniques.\cite{7}  Optical SHG\cite{15} involves the conversion of light (electric field $E^\omega$) at a frequency $\omega$ into an optical signal at a frequency 2$\omega$ by a nonlinear medium, through the creation of a nonlinear polarization $P_i^{2\omega}\propto d_{ijk}E_j^{\omega}E_k^{\omega}$, where $d_{ijk}$ represent the non-linear optical coefficients. BFO film with thickness of about 50 nm grown on STO(111) substrates was used for this study. STO  is centrosymmetric (cubic) and does not contribute SHG signals of its own for the incident powers used. The SHG experiment was performed with a fundamental wave generated from a tunable Ti-sapphire laser with 65 fs pulses of wavelength 800 nm incident from the substrate side at variable tilt angles $\theta$ to the sample surface normal. 

As shown in  Fig. \ref{fig:2}, the crystallographic \textit{y}-\textit{z} plane in the BFO film  was aligned with the incidence plane. The polarization direction of incident light is at an angle $\phi$ from the \textit{x} axis, which was rotated continuously using a half-wave plate . The intensity $I_j^{2\omega}$ of the output SHG signal at 400 nm wavelength from the film was detected along either $j=p,s$ polarization directions as a function of polarization angle $\phi$ of incident light. The resulting polar plots of SHG intensity for $p$ and $s$-polarized output at $\theta = 0^\circ$ and $\theta = 45^\circ$ are shown in Fig. \ref{fig:2}(a) and (b) respectively. If the incident beam has intensity $I_0$ then the nonlinear polarizations for BFO(111) film with \textit{x}-axis perpendicular to the plane of incidence is given by\cite{16,17}, 
\begin{eqnarray}
 \label{eq:1}
P_x^{NL} =&& I_0f_x \sin 2\phi (d_{15}f_z \sin\theta - d_{22}f_y \cos\theta ) \nonumber \\
P_y^{NL} =&& I_0(- d_{22}\cos^2\phi f_x^2 + d_{22}f_y^2\cos^2\theta \sin^2 \phi \nonumber \\
+&& d_{15}f_y f_z\sin 2\theta \sin^2 \phi) \nonumber \\
P_z^{NL} =&& I_0(- d_{31}\cos^2\phi f_x^2 + d_{31}f_y^2\cos^2\theta \sin^2 \phi \nonumber\\
+&& d_{33}f_z^2\sin^2 \theta \sin^2 \phi) 
\end{eqnarray} where $d_{ij}$ are nonlinear coefficients and $f_i$ are effective linear Fresnel coefficients. The measured intensity of the \textit{p} and \textit{s} polarized SHG in transmission geometry (neglecting birefringence) is proportional to nonlinear polarization. The expected SHG intensity expressions for $p$ and $s$ output polarizations in the predicted 3$m$ symmetry system of BFO are:
\begin{eqnarray}
\label{eq:2}
 I_p^{2\omega} = && A( \cos^2 \phi+ B \sin^2 \phi)^2 \nonumber \\
 I_s^{2\omega} = && C \sin^2 2\phi
\end{eqnarray}
where B and C are given by
\begin{widetext}
\begin{eqnarray}
\label{eq:3}
B =&& \frac{K_{15}f_y f_z\sin2\theta \cos\theta_B+K_{31}f_y^2\cos^2\theta\sin\theta_B+K_{33}f_z^2\sin^2\theta 
+ f_y^2\cos^2\theta\cos\theta_B} {K_{31}f_x^2\sin\theta_B-f_x^2\cos\theta_B}\nonumber\\
C =&& D(K_{15}f_z f_x\sin\theta-f_x f_y\cos\theta)
\end{eqnarray}
\end{widetext} 

Here $K_{15}=d_{15}/d_{22}$, $K_{31}=d_{31}/d_{22}$ and $K_{33}=d_{33}/d_{22}$ are the ratios of the nonlinear optical coefficients, A and D are scaling parameters, and $\theta_B$ is the angle that the generated second harmonic wave makes with the surface normal \textit{inside} the film. Theoretical fits to the experimental polar plots based on Eq. \ref{eq:2} are excellent both in normal incidence  and tilted configuration as shown in Fig. \ref{fig:2}(a) and  (b), respectively, for both \textit{p} and \textit{s} polarized SHG output.
\begin{figure}
 \centering
 \includegraphics[width=0.47 \textwidth,,height=0.22 \textheight,angle=0]{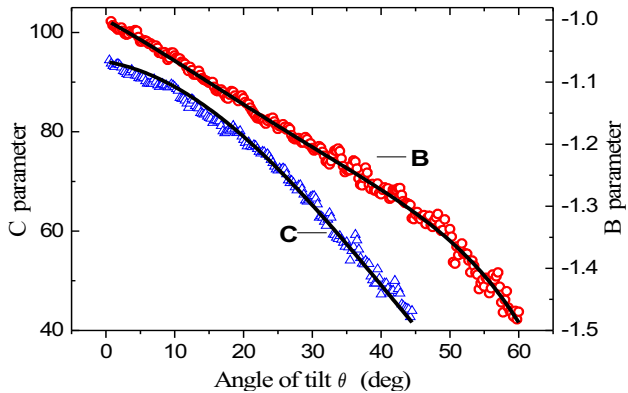}
\caption{\label{fig:3}Variation of extracted B parameter from $p$ polarized SHG signal ($B_{expt} = - [{\frac{I_p^{2\omega}(\phi=90^\circ)}{I_p^{2\omega}(\phi=0^\circ)}}]^{\frac{1}{2}}$) and C Parameter ($C_{expt} = I_s^{2\omega}(\phi=45^\circ)$) as a function of tilt angle, $\theta$. The solid lines show theoretical fits to the data.}
\end{figure}
In normal incidence ($\theta=0^\circ$), only the $d_{22}$ coefficient is involved in the generated $I_p^{2\omega}(\phi=0^\circ)$. Using the $d_{22}$=1.672 pm/V coefficient of a single crystal \textit{z}-cut LiTaO$_3$ used as a reference, the d$_{22}$ coefficient of the BFO film is calculated by employing the following equation\cite{18},
\begin{equation}
\label{eq:4}
\frac{d_{eff}^2}{d_r^2}=\frac{P_f^{2\omega}A_f n_f^{2\omega}}{P_r^{2\omega}A_r n_r^{2\omega}}\Big(\frac{P_r^\omega T_r^\omega n_f^\omega}{P_f^\omega T_f^\omega n_r^\omega}\Big)^2 \frac{l_{c,f}}{l_{c,r}} \frac{\int\limits_0^{l_{c,r}}x^2 e^{-\alpha_r^{2\omega}(l_{c,r}-x)} \,dx}{\int\limits_0^{l_{c,f}}x^2 e^{-\alpha_f^{2\omega}(l_{c,f}-x)}\,dx}, 
\end{equation}where the subscripts $\textit{r}$ and $\textit{f}$  refer to the reference and the film, respectively, \textit{$P^{2\omega}$}(\textit{$P^\omega$}) are the second-harmonic (fundamental) signal powers measured, T is the transmission coefficient of the fundamental, A is the area of the probed beam, $\textit{n}$ are the indices of refraction, \textit{$l_c$} the coherence lengths and $\alpha = 4\pi$k$/\lambda$ is the absorption coefficient  at 2$\omega$. For this experiment, $A_f =A_r=\pi(60 \mu$m$)^2$ and the coherence length of the film $l_c$(film) is equal to  the film thickness.

 The B and C parameters, which contain the ratios of nonlinear coefficients, are experimentally obtained by collecting the \textit{p}-in-\textit{p}-out $I_p^{2\omega}(\phi=90^\circ)$, \textit{s}-in-\textit{p}-out $I_p^{2\omega}(\phi=0^\circ)$ and 45-in-\textit{s}-out $I_s^{2\omega}(\phi=45^\circ)$ SHG signals for different angles of tilt $\theta$ about the \textit{x} axis. The experimental data for B and C parameters (Fig. \ref{fig:3}) is then fitted to Eq. \ref{eq:3} to extract the ratios $K_{15} = 0.20\pm0.01$, $K_{31} = 0.35\pm0.02$ and $K_{33} = -11.4\pm0.20$. Taking absorption into account, the estimated effective coefficients are \begin{eqnarray*}
\left|d_{22}\right| =&& 298.4\pm6.1 \; \textrm{pm/V},\; \left|d_{31}\right| = 59.7\pm4.2 \; \textrm{pm/V}\\
\left|d_{15}\right| =&& 104.4\pm8.1 \; \textrm{pm/V},\; \left|d_{33}\right| = 3401\pm129 \; \textrm{pm/V}.
\end{eqnarray*}
Note that only the signs of the ratios $K_{15},K_{31}$ and $K_{33}$ were determined unambiguously.  The absolute signs of the $d_{ij}$ coefficients were not determined, except to state that the $d_{33}$ coefficient has the opposite sign to the other coefficients. The large values of $d_{ij}$ coefficients most likely arise due to electronic resonances at the 400 nm SHG wavelength.

To conclude, we report the complex index of refraction versus wavelength and optical second harmonic generation coefficients in BiFeO$_3$ thin films.  These studies will be important in performing further linear and nonlinear optical spectroscopy of the magnetism and ferroelectricity in this material.

We would like to acknowledge support from NSF under grant Nos. DMR-0507146, DMR-0512165, DMR-0602986 and DMR-0213623. At Univ. of Tennessee, the research was supported by the Materials Science Division, BES, U.S. DoE (DE-FG02-01ER45885).

\end{document}